\documentclass{sigchi}

\CopyrightYear{2020}
\setcopyright{acmlicensed}

\doi{https://doi.org/10.1145/3313831.3376533}
\isbn{978-1-4503-6708-0/20/04}
\conferenceinfo{CHI'20,}{April 25--30, 2020, Honolulu, HI, USA}
\acmPrice{\$15.00}

\usepackage{balance}       
\usepackage{graphics}      
\usepackage[T1]{fontenc}   
\usepackage{txfonts}
\usepackage{mathptmx}
\usepackage[pdflang={en-US},pdftex]{hyperref}
\usepackage{color}
\usepackage{booktabs}
\usepackage{textcomp}
\usepackage{tabularx}
\usepackage{booktabs}
\usepackage{multirow}

\usepackage{subcaption}
\usepackage{enumitem}
\setitemize{noitemsep,topsep=5pt,parsep=0pt,partopsep=0pt}
\usepackage{xspace}
\usepackage{xstring}

\usepackage{microtype}        
\usepackage{ccicons}          

\usepackage{todonotes}

\def\plaintitle{Paths Explored, Paths Omitted, Paths Obscured: Decision Points \& Selective Reporting in End-to-End Data Analysis}
\def\plainauthor{Yang Liu, Tim Althoff, Jeffrey Heer}
\def\plainkeywords{Data analysis; Analytic decision making; Multiverse analysis; Garden of forking paths; Reproducibility; Interview Study}

\makeatletter
\def\url@leostyle{%
  \@ifundefined{selectfont}{
    \def\UrlFont{\sf}
  }{
    \def\UrlFont{\small\bf\ttfamily}
  }}
\makeatother
\def\pprw{8.5in}
\def\pprh{11in}

\definecolor{linkColor}{RGB}{6,125,233}
\hypersetup{%
  pdftitle={\plaintitle},
  pdfauthor={\plainauthor},
  pdfkeywords={\plainkeywords},
  pdfdisplaydoctitle=true, 
  bookmarksnumbered,
  pdfstartview={FitH},
  colorlinks,
  citecolor=black,
  filecolor=black,
  linkcolor=black,
  urlcolor=linkColor,
  breaklinks=true,
  hypertexnames=false
}

\newcommand{\rfig}[1]{Figure~\ref{#1}}
\newcommand{\qi}[1]{\textit{``#1''}}

\newcommand{\ie}{{\em i.e.,}\xspace}
\newcommand{\eg}{{\em e.g.,}\xspace}
\newcommand{\cf}{{\em c.f.}\xspace}
\newcommand{\etal}{{\em et~al.}\xspace}

\renewenvironment{quote}{%
   \list{}{%
   	 \vspace{-3pt}
     \leftmargin0.34cm   
     \rightmargin\leftmargin
   }
   \item\relax
}
{\vspace{-3pt}\endlist}
\newcommand{\q}[1]{\begin{quote}\emph{``#1''}\end{quote}}

\newcommand{\glyph}[1]{%
  \begingroup\normalfont
  \includegraphics[height=\fontcharht\font`\B]{#1.png}%
  \endgroup
}
\newcommand{\legend}[1]{(\glyph{#1})}

\newcommand{\seefig}[2]{\IfEq{#1}{1}{(\rfig{fig:p1}, {#2})}{(\rfig{fig:diagrams}{#1}, {#2})}}

\begin{document}

\title{\plaintitle}

\numberofauthors{3}
\author{%
  \alignauthor{Yang Liu\\
    \affaddr{University of Washington}\\
    \email{yliu0@uw.edu}\\}
  \alignauthor{Tim Althoff\\
    \affaddr{University of Washington}\\
  	\email{althoff@cs.washington.edu}}
  \alignauthor{Jeffrey Heer\\
    \affaddr{University of Washington}\\
  	\email{jheer@uw.edu}}
}

\maketitle

\begin{abstract}
Drawing reliable inferences from data involves many, sometimes arbitrary, decisions across phases of data collection, wrangling, and modeling.
As different choices can lead to diverging conclusions, understanding how researchers make analytic decisions is important for supporting robust and replicable analysis.
In this study, we pore over nine published research studies and conduct semi-structured interviews with their authors.
We observe that researchers often base their decisions on methodological or theoretical concerns, but subject to constraints arising from the data, expertise, or perceived interpretability.
We confirm that researchers may experiment with choices in search of desirable results, but also identify other reasons why researchers explore alternatives yet omit findings.
In concert with our interviews, we also contribute visualizations for communicating decision processes throughout an analysis.
Based on our results, we identify design opportunities for strengthening end-to-end analysis, for instance via tracking and meta-analysis of multiple decision paths.

\end{abstract}




\setlength{\defaultaddspace}{2pt} 

\newcommand{\tableCodes}{
\begin{table*}[h]
\small
\begin{tabularx}{\textwidth}{llXp{.3\textwidth}r}
\textbf{Theme} & \textbf{Category} & \textbf{Description} & \textbf{Representative Quote} & \textbf{\%} \\ \toprule
\multirow{6}{4em}[-2.7em]{Decision rationales} & Methodology & Participants defend the decision with methodological concerns, including statistical validity, study design and research scope. & I mainly used t-test for hypothesis testing because my data was parametric. & 25 \\ \addlinespace 
& Prior work & Participants support the analytic decision using previous studies, ``standard practice'' and/or internalized knowledge. & We adapt the method from a previous paper and we follow the same process to do the analysis. & 33 \\ \addlinespace 
& Data & Participants mention data constraints, including data availability, data size and data quality. & The reason I combined them together is because more data has less variation. & 21 \\ \addlinespace 
& Expertise & Participants feel limited by expertise. & I don't know how to do this really. & 12 \\ \addlinespace 
& Communication & Participants prefer an alternative that is easier to communicate. & Because they were actually very hard to write up. & 7 \\ \addlinespace 
& Sensitivity & Participants believe that the decision has little impact on the results and provide no further rationales. & In my quick mental calculation, it seemed like it wouldn't actually make a big difference. & 3 \\ \hline \addlinespace 
\multirow{4}{4em}[-2.2em]{Executing alternatives} & Opportunism & Participants willingly explore new alternatives to look for desired results. & I tried three different settings for those parameters and the chosen ones looked slightly better. & 45 \\ \addlinespace 
& Systematicity & Participants outline all reasonable alternatives, implement them, and choose the winning alternative based on an objective metric. & We performed a sensitivity analysis to identify the best combination. & 9 \\ \addlinespace 
& Robustness & Participants implement additional alternatives after making a decision, in order to gauge the robustness of their conclusions. & That is just for robustness, to say, "even if you look at [another option], you see the same thing." & 16 \\ \addlinespace 
& Contingency & Participants have to deviate from their original plans because the planned analysis turned out to be erroneous and/or infeasible. & This [filter] produced anomalous results and we went back [to apply] a more stringent filtering. & 30 \\ \hline \addlinespace 
\multirow{4}{4em}[-2.2em]{Selective reporting} & Desired results & Participants only report the desired results and omit findings that are non-significant, uninteresting, or incoherent to their theory. & It felt stronger to say five out of seven, rather than four out of six, was one reason to keep it. & 29 \\ \addlinespace 
& Similar results & Participants claim that the results are similar and thus omit interchangeable alternative analyses. & But it didn't make a huge difference so we just kind of went with [the current option]. & 10 \\ \addlinespace
& Correctness & Participants apply rationales, primarily methodology and prior work, to remove analytic approaches they consider incorrect. & I was concerned about whether I had a strong hypothesis to see those interaction effects or not. & 31 \\ \addlinespace
& Social constraints & Social constraints and communication concerns prevent participants from reporting some findings. & I'm a second author and many decisions made in the manuscript writing were against my wishes. & 31 \\ \addlinespace
\bottomrule
\end{tabularx}
\vspace{-8pt}
\caption{Themes and high-level codes that emerged from open coding of the interview data.}\vspace{-10pt}
\label{table:code}
\end{table*}
}

\newcommand{\figureChecklist}{
\begin{figure*}[h]
    \centering
	\includegraphics[width=1.6\columnwidth,trim={4cm 7cm 4cm 2cm}, clip]{checklist.pdf}
	\vspace{-7pt}
    \caption{A checklist of decision points we shared with the participants during phase 2 of the interviews.}
    \label{fig:checklist}
\end{figure*}
}

\newcommand{\figurePOne}{
\begin{figure*}[htb]
	\vspace{-22pt}
	\centering
	\includegraphics[width=1.54\columnwidth,trim={0 10cm 25cm 0}, clip]{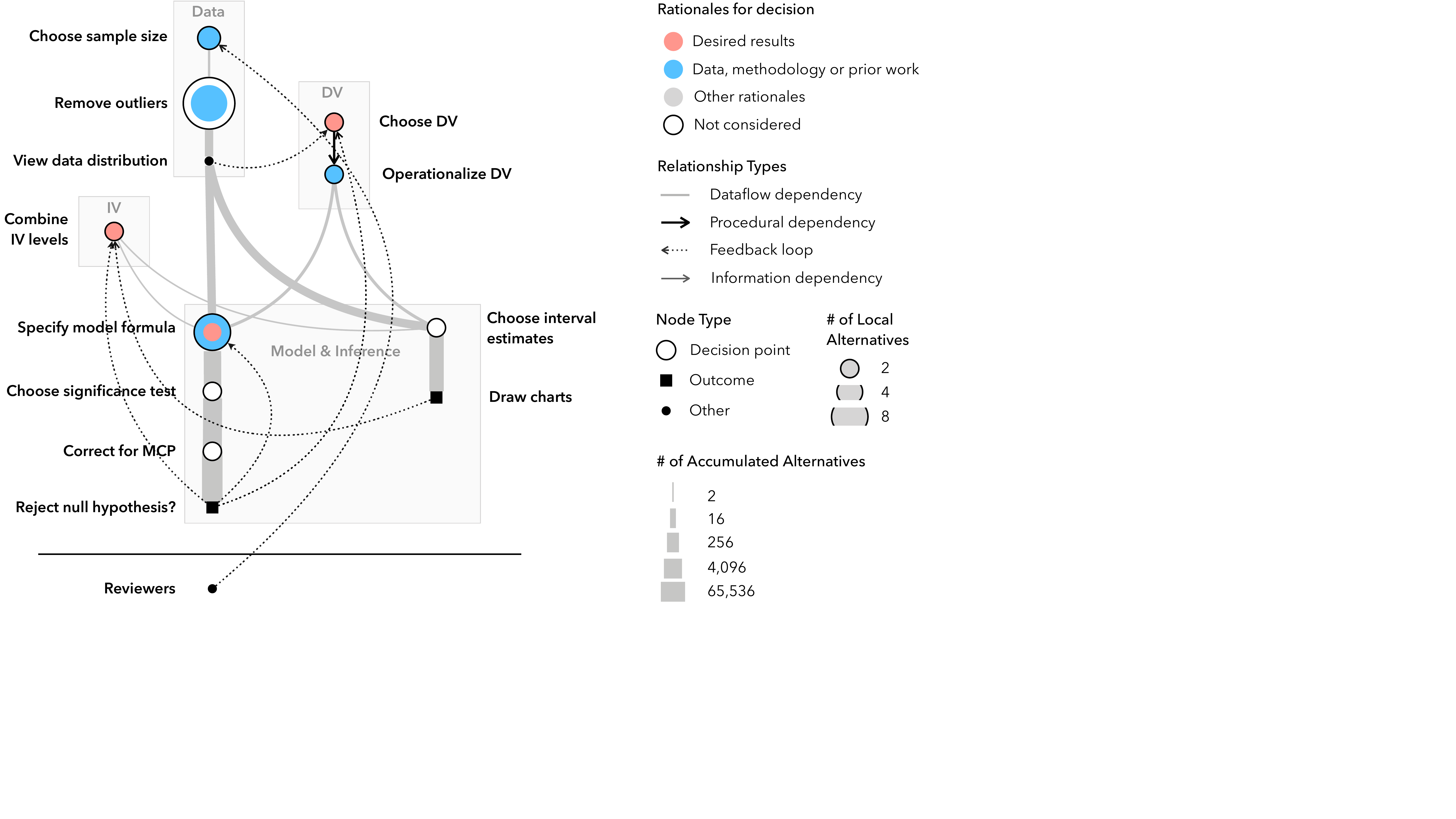}
	\vspace{-10pt}
    \caption{Analytic Decision Graph for P1, representing a controlled experiment to investigate the impact of web design on reading performance. At several steps, P1 revised her analytic decisions based on end results and reviewer feedback, for instance merging two levels of an IV because effect sizes were similar. While she examined model specification options thoroughly, she appeared to place less emphasis on inference decisions such as choosing which significance test to use.}
    \vspace{-10pt}
    \label{fig:p1}
\end{figure*}
}

\newcommand{\figureDiagrams}{
\begin{figure*}
	\centering
	\begin{subfigure}[t]{0.32\textwidth}
        \centering
		\includegraphics[width=1\columnwidth,trim={0.8cm 2.8cm 28cm 1cm}, clip]{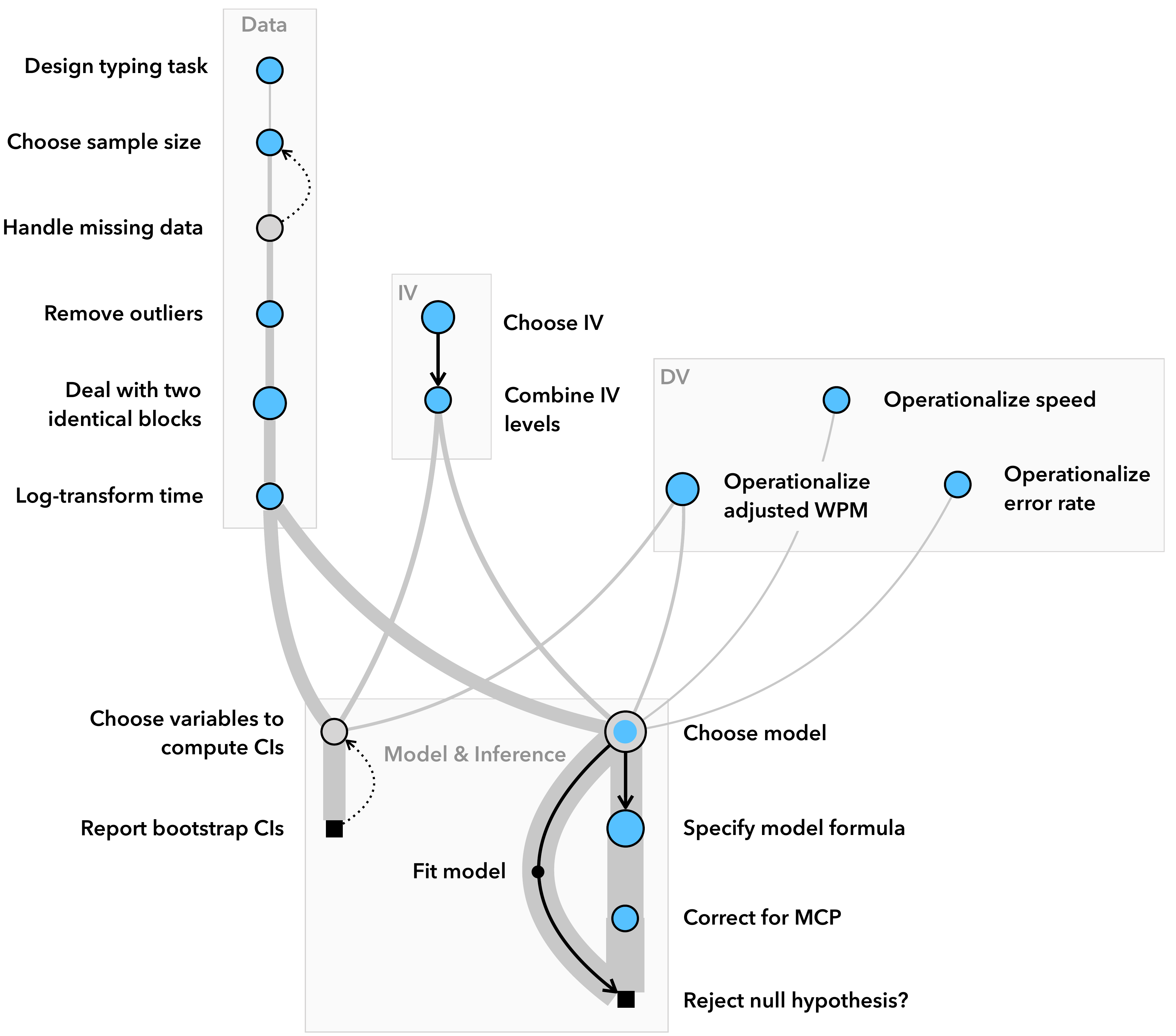}
        \caption{P2}
    \end{subfigure}%
    ~ 
    \begin{subfigure}[t]{0.32\textwidth}
        \centering
        \includegraphics[width=1\columnwidth,trim={1.5cm 1.5cm 28cm 1.9cm}, clip]{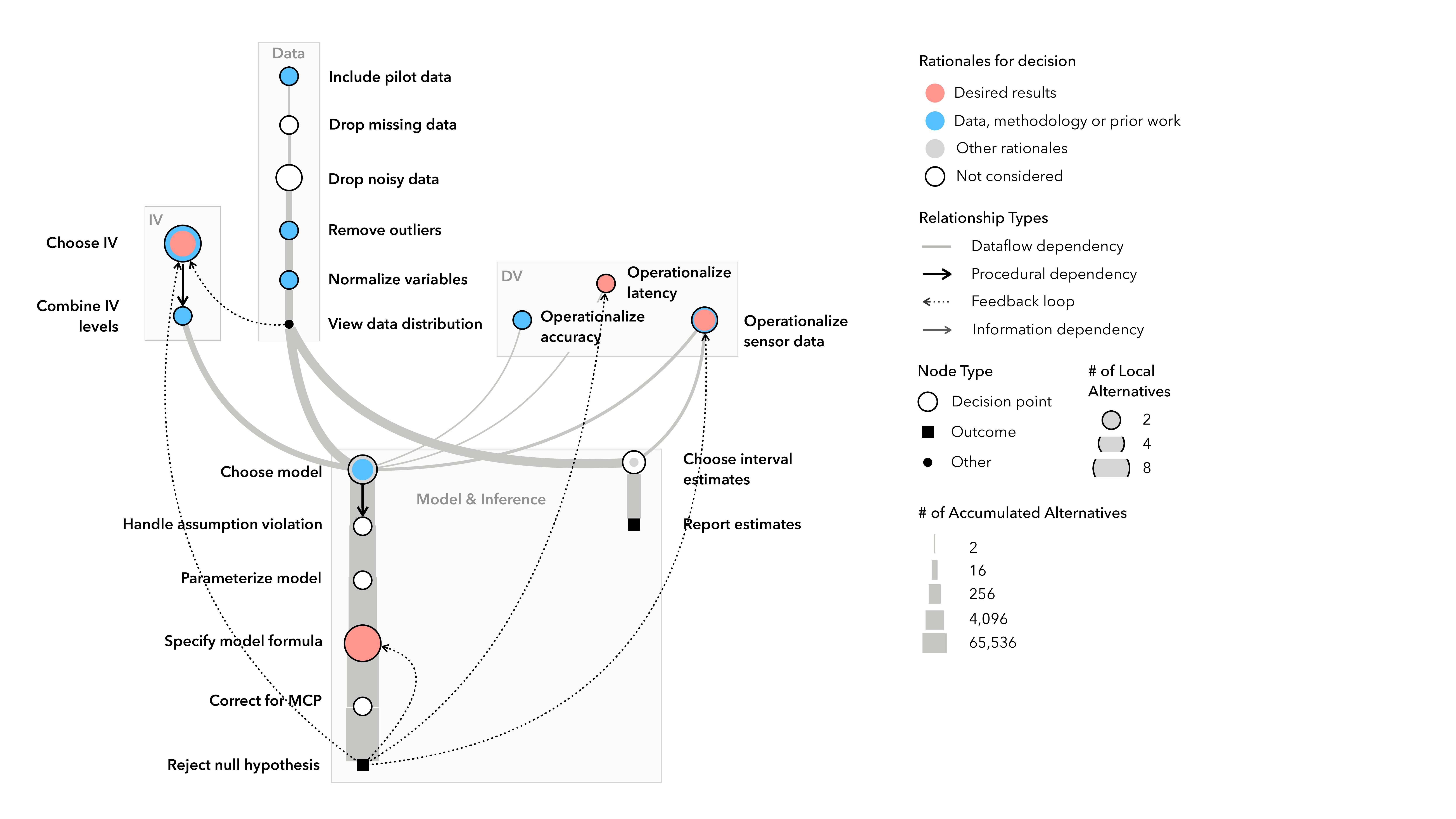}
        \caption{P3}
    \end{subfigure}%
    ~
	\begin{subfigure}[t]{0.32\textwidth}
        \centering
		\includegraphics[width=1\columnwidth,trim={0cm 7cm 31cm 1cm}, clip]{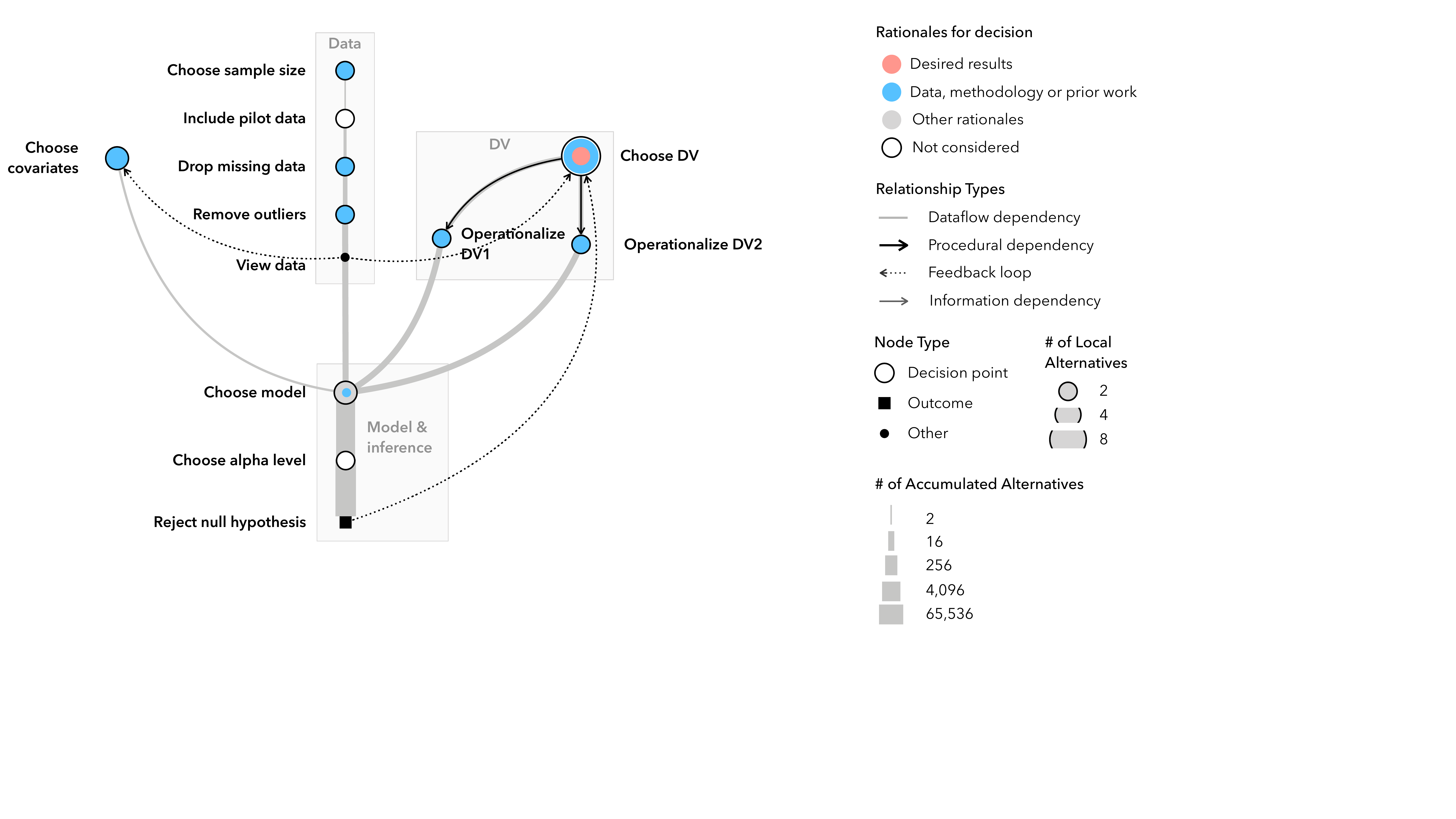}
        \caption{P4}
    \end{subfigure}
    \begin{subfigure}[t]{0.28\textwidth}
        \centering
        \includegraphics[width=1\columnwidth,trim={0cm 3cm 35cm 0cm}, clip]{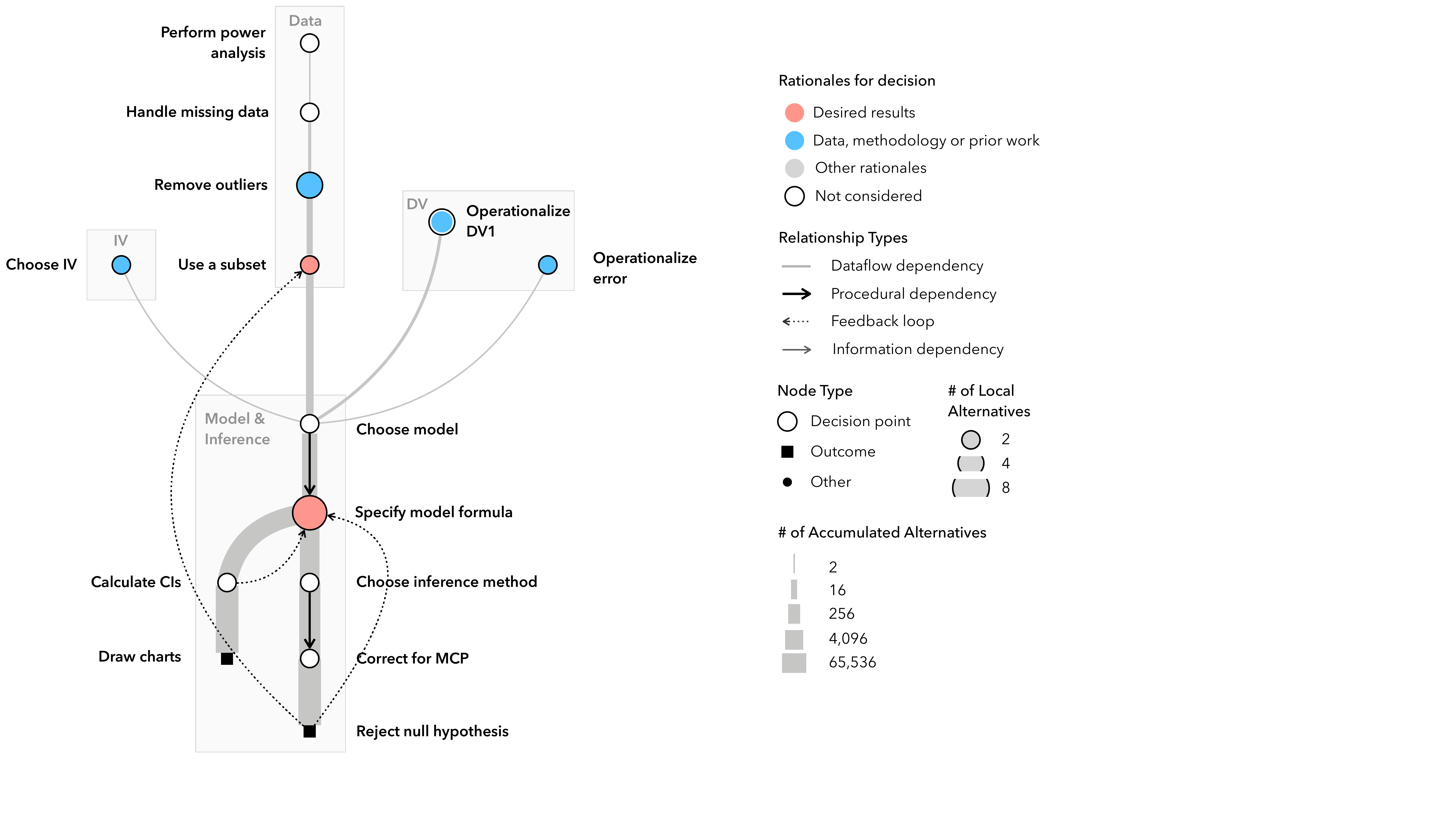}
        \caption{P5}
    \end{subfigure}%
    ~
	\begin{subfigure}[t]{0.32\textwidth}
        \centering
		\includegraphics[width=1\columnwidth,trim={3cm 2cm 28cm 0.5cm}, clip]{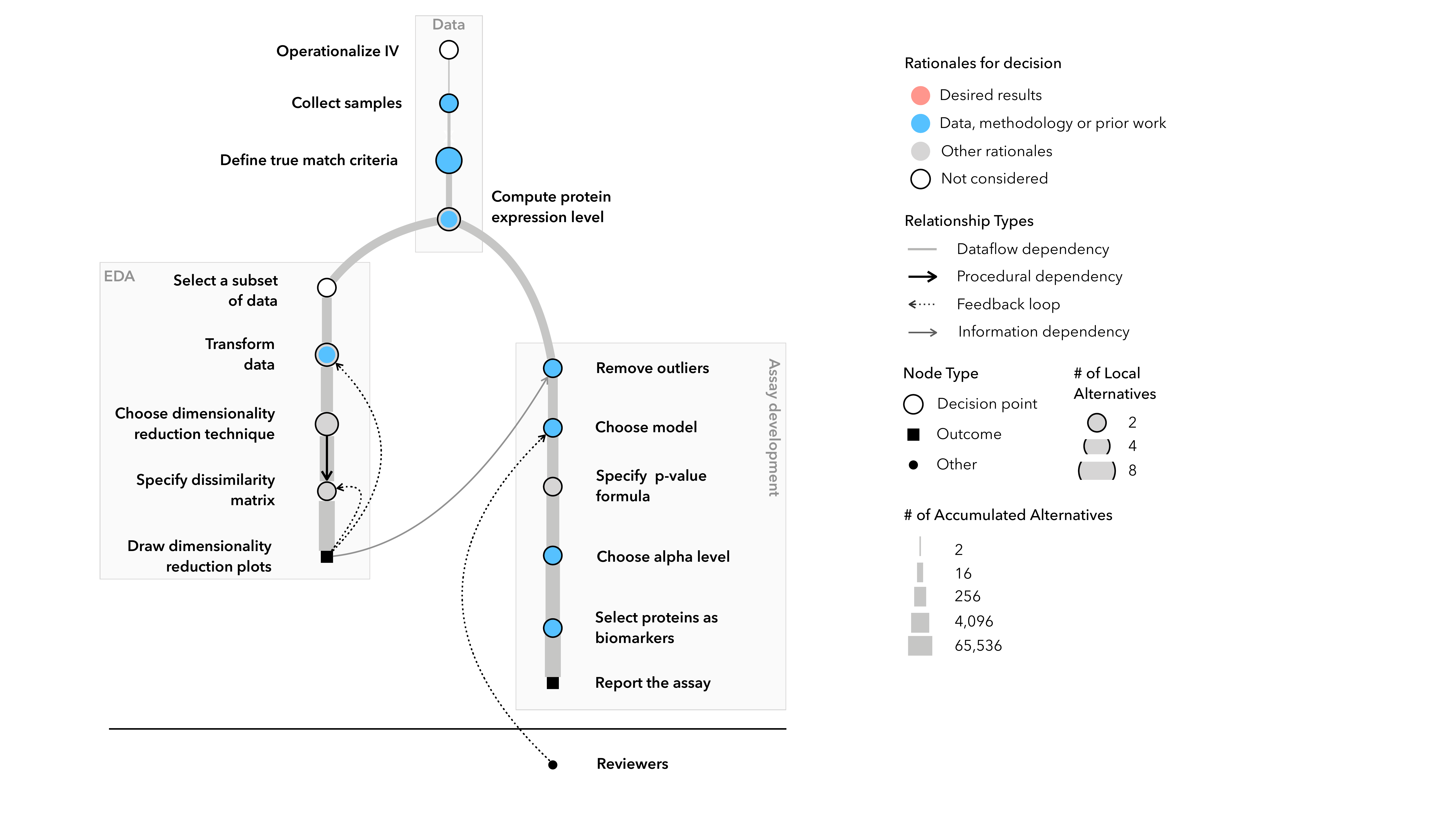}
        \caption{P6}
    \end{subfigure}%
    ~ 
    \begin{subfigure}[t]{0.4\textwidth}
        \centering
        \includegraphics[width=1\columnwidth,trim={1.5cm 6cm 26cm 0.5cm}, clip]{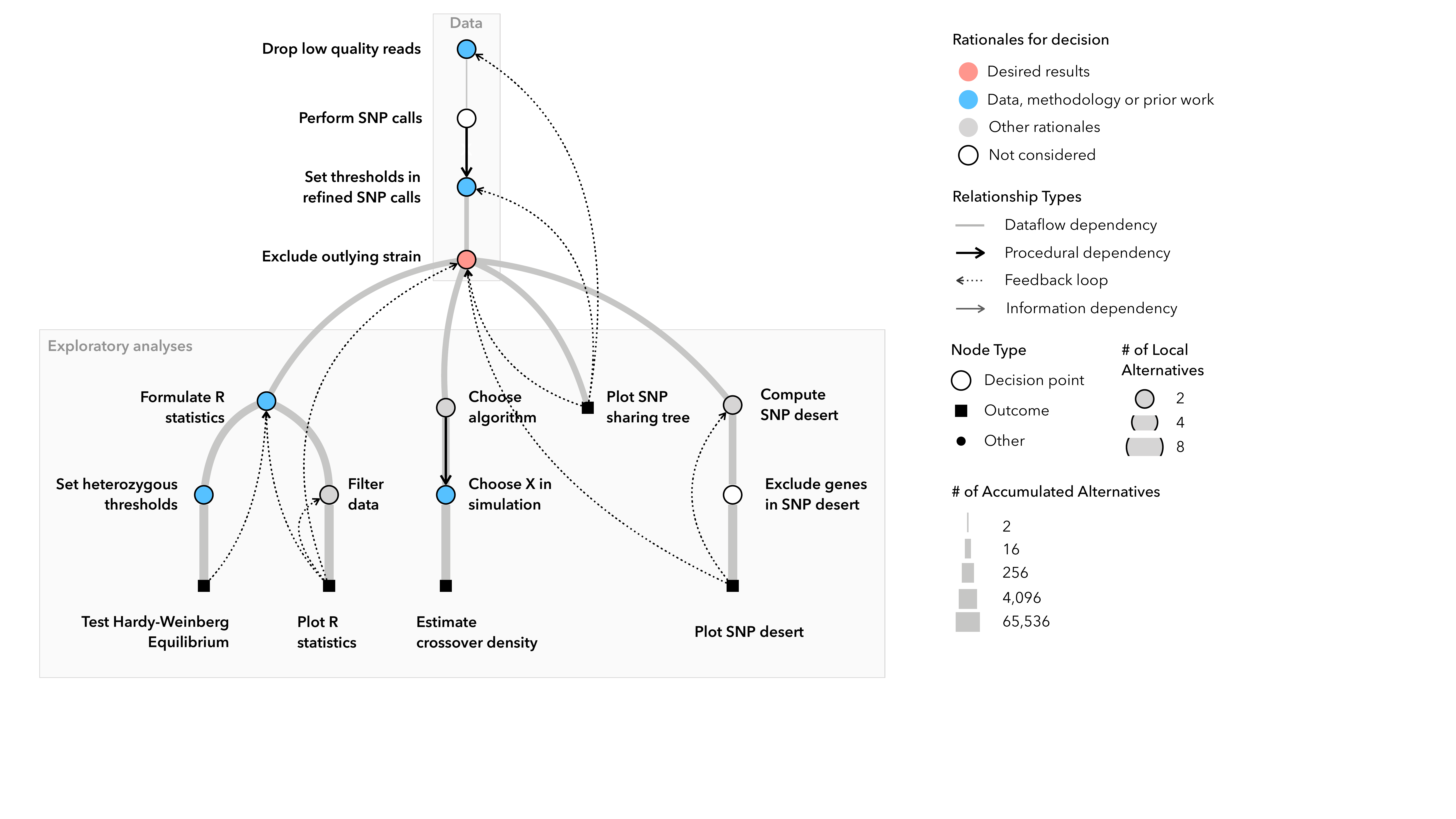}
        \caption{P7}
    \end{subfigure}
    \begin{subfigure}[t]{0.28\textwidth}
        \centering
        \includegraphics[width=0.5\columnwidth,trim={30cm 7cm 25cm 0cm}, clip]{p1.pdf}
        \caption{Legends}
    \end{subfigure}%
    ~
	\begin{subfigure}[t]{0.32\textwidth}
        \centering
		\includegraphics[width=1\columnwidth,trim={1cm 0cm 30cm 0cm}, clip]{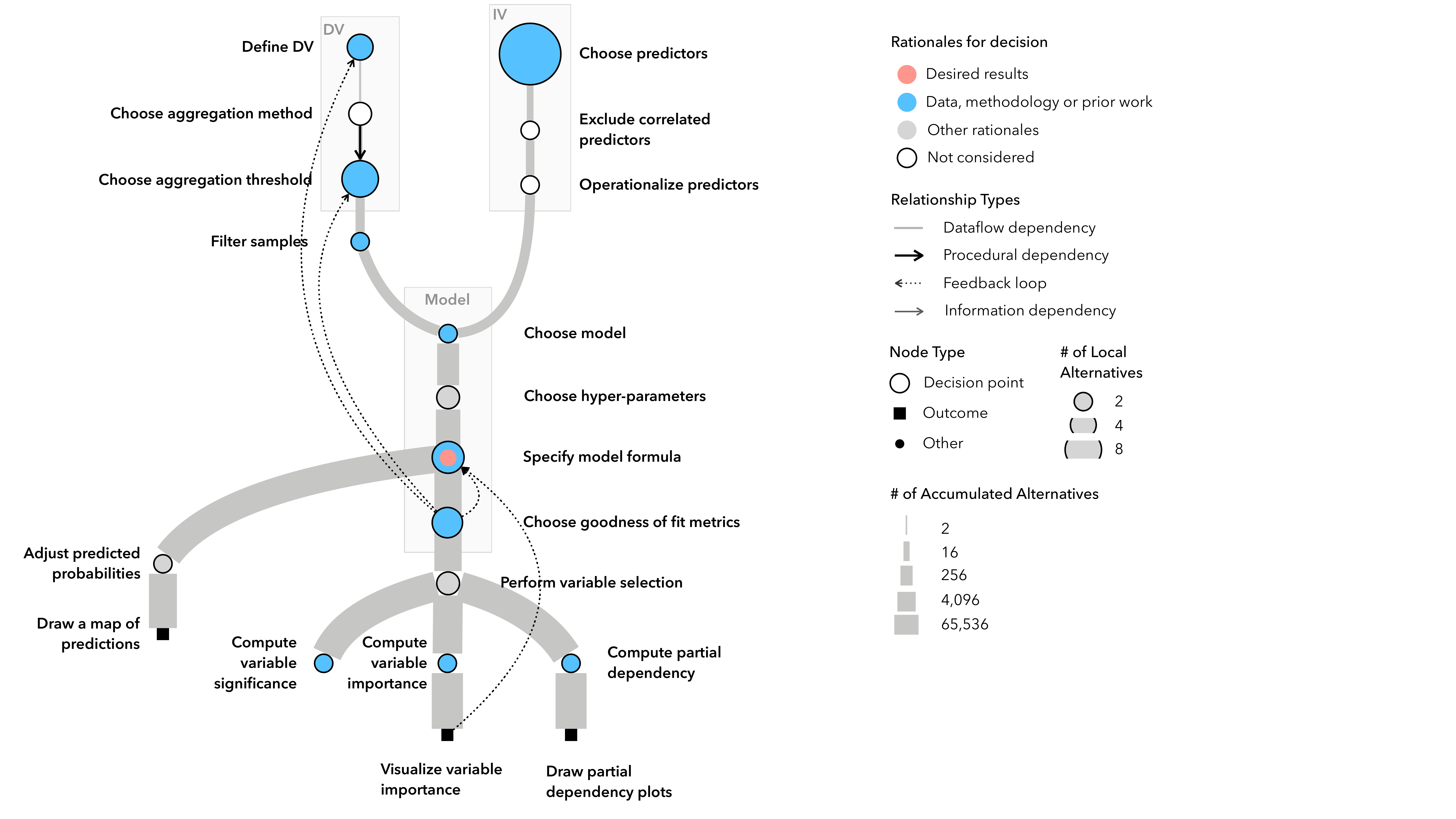}
        \caption{P8}
    \end{subfigure}%
    ~
    \begin{subfigure}[t]{0.4\textwidth}
        \centering
        \includegraphics[width=1\columnwidth,trim={1.5cm 0cm 26cm 0cm}, clip]{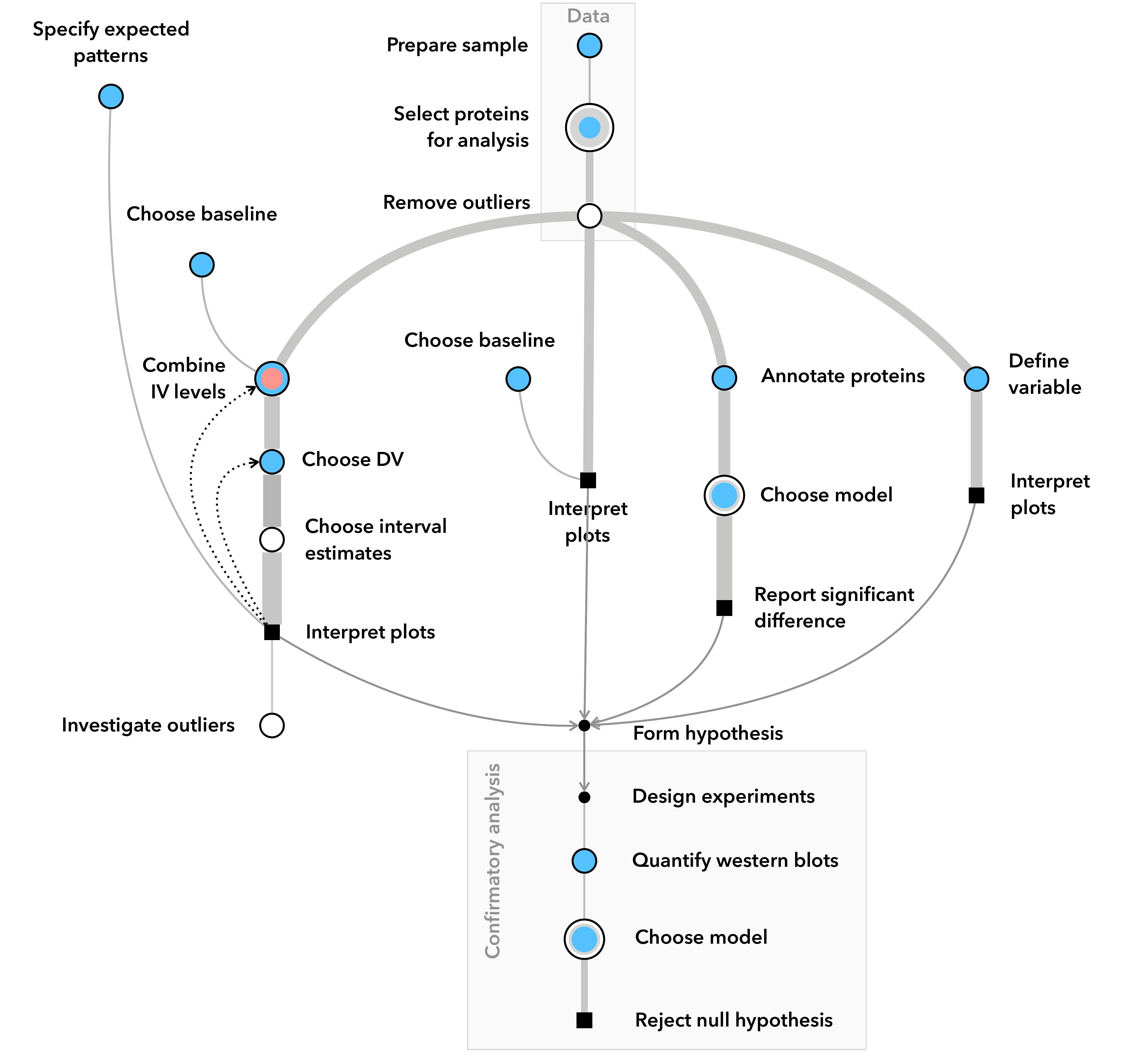}
        \caption{P9}
    \end{subfigure}
    \caption{Analytic Decision Graphs for P2--P9.}
    \label{fig:diagrams}
\end{figure*}
}

\newcommand{\diagramOne}{
\begin{figure*}[htb]
	\vspace{-7pt}
	\centering
	\includegraphics[width=1.55\columnwidth,trim={0 9cm 25cm 0}, clip]{p1.pdf}
	\vspace{-7pt}
    \caption{P1 designed a controlled experiment to investigate the impact of web design on reading performance. At several steps, she revised her analytic decisions based on end results and reviewer feedback, for instance merging two levels of an IV because effect sizes were similar. While she examined model specification options thoroughly, she appeared to have placed less emphasis on inference decisions.}
    \label{fig:diagram_one}
\end{figure*}
}

\newcommand{\diagramTwo}{
\begin{figure*}[htb]
	\centering
	\includegraphics[width=1.8\columnwidth,trim={0.8cm 2.8cm 12.8cm 1cm}, clip]{p2.pdf}
	\vspace{-7pt}
    \caption{P2 used controlled experiments to evaluate a throughput metric for text entry. He defended most of his analytic decisions using methodological concerns, prior work and data constraints. His analysis process included two feedback loops, where he went back to data collection to gather missing data, and omitted a non-significant result.}
\end{figure*}
}

\newcommand{\diagramThree}{
\begin{figure*}[htb]
	\centering
	\includegraphics[width=1.7\columnwidth,trim={1.5cm 1.5cm 12.8cm 1.9cm}, clip]{p3.pdf}
	\vspace{-7pt}
    \caption{P3 analyzed smartphone data from controlled experiments to test several hypotheses. He experimented with various analytic paths to obtain significant p-values, including multiple ways to operationalize DVs, different IVs, and different model formula. Apart from model choice and model specification, he seemed to have overlooked other modeling and inference decisions, for example neglecting multiple comparison problem and ignoring model assumptions.}
\end{figure*}
}

\newcommand{\diagramFour}{
\begin{figure*}[htb]
	\centering
	\includegraphics[width=1.8\columnwidth,trim={0cm 7cm 13cm 1cm}, clip]{p4.pdf}
	\vspace{-7pt}
    \caption{P4 compared their smartphone notification system to the baseline in a simple within-subject experiment. Since they analyzed raw logs from smartphones, many potential variables were reasonable proxies for performance, making ``choosing DV'' the largest decision point. They finally chose the variables that had good data quality and yielded significant p-values.}
\end{figure*}
}

\newcommand{\diagramFive}{
\begin{figure*}[htb]
	\vspace{-15pt}
	\centering
	\includegraphics[width=1.6\columnwidth,trim={0cm 3cm 19cm 0cm}, clip]{p5.pdf}
	\vspace{-7pt}
    \caption{P5 designed controlled experiments to evaluate several interaction techniques. As it was an earlier project, P5 admitted being unfamiliar with some techniques (\eg power analysis and Bayesian models) and thus neglected some analytic options. She made two analytic decisions after seeing the results: whether to use a subset of the data, and which model specification to adopt.}
\end{figure*}
}

\newcommand{\diagramSix}{
\begin{figure*}[htb]
	\vspace{-7pt}
	\centering
	\includegraphics[width=1.6\columnwidth,trim={3cm 2cm 13cm 0.5cm}, clip]{p6.pdf}
	\vspace{-7pt}
    \caption{P6 analyzed proteomic profiles of geoduck clams in different reproductive maturation stages to develop an assay for maturation status. She first performed exploratory analysis to visualize the samples in nonmetric multidimensional scaling (NMDS) plots and then selected biomarker proteins to include in the assay. She iterated on the NMDS plots, but the analytic decisions did not change the final result (\ie the assay).}
\end{figure*}
}

\newcommand{\diagramSeven}{
\begin{figure*}[htb]
	\vspace{-7pt}
	\centering
	\includegraphics[width=1.7\columnwidth,trim={1.5cm 6cm 10cm 0.5cm}, clip]{p7.pdf}
	\vspace{-7pt}
    \caption{P7 performed a suite of exploratory analyses on the genomic data of ocean algae and revealed evidence on asexual lineage. The analytic process branched into multiple investigations, each leading to a separate piece of evidence to support the final conclusion. P7 frequently revised analytic decisions to discern patterns from noise, as is indicated by the numerous feedback loops throughout the graph.}
\end{figure*}
}

\newcommand{\diagramNine}{
\begin{figure*}[htb]
	\vspace{-7pt}
	\centering
	\includegraphics[width=1.9\columnwidth,trim={1.5cm 0cm 9.5cm 0cm}, clip]{p8.pdf}
	\vspace{-7pt}
    \caption{To investigate the effect of a gene, P9 performed exploratory analyses on proteomic data of fruit fly mutants, formed a hypothesis, and directly tested the hypothesis in a subsequent experiment. Her analysis process expanded into multiple branches during the exploratory phase and narrowed down into a single confirmatory conclusion.}
    \label{fig:diagram_nine}
\end{figure*}
}

\newcommand{\diagramEight}{
\begin{figure*}[htb]
	\vspace{-10pt}
	\centering
	\includegraphics[width=1.9\columnwidth,trim={1cm 0cm 12cm 0cm}, clip]{p9.pdf}
	\vspace{-10pt}
    \caption{P8 built a machine learning model to draw exploratory conclusions on what might affect agricultural land abandonment in Europe. They used a local ``multiverse'' approach, where they took all combinations of 3 abandonment definitions and 8 aggregation thresholds, and used the goodness of fit of the model to identify the best combination. They also selected the best model specification based on goodness of fit, but later modified the model to remove two dominant predictors. The analysis process was fully disclosed in the research report, including all unsuccessful analytic paths. }
\end{figure*}
}

\newcommand{\diagramOurs}{
\begin{figure*}
	\vspace{-10pt}
	\centering
	\includegraphics[width=1.9\columnwidth,trim={0cm 10cm 18cm 0cm}, clip]{ours.pdf}
	\vspace{-5pt}
    \caption{ADG of our own analysis. This is a proof-of-concept that ADGs might be applied to qualitative data analysis.}
    \label{fig:diagram_ours}
\end{figure*}
}

\section{Introduction}

A replicability crisis has stirred multiple scientific fields~\cite{baker2016}, with replication studies failing to validate prior results~\cite{begley2012, border2019, open2015, prinz2011}.
In Biology, two laboratories ventured to validate published ``landmark'' studies, but were successful in replicating the original results in only 11\% and 25\% of projects, respectively~\cite{begley2012, prinz2011}.
In Psychology, the Open Science Collaboration replicated 100 published studies using high-powered designs and original materials, but found that on average,  ``replication effects were half the magnitude of original effects''~\cite{open2015}.

Why are these studies, backed by empirical evidence from peer-reviewed data analysis, failing to replicate?
Scholars suggest that undisclosed freedom in analytic decisions plays a key role~\cite{gelman2013garden, simmons2011}.
Researchers routinely make decisions throughout quantitative data analysis, from data collection and wrangling, to statistical modeling and inference.
For example, what are the cutoffs for excluding outliers? What variations of model formulae should one choose?
Different sequences of analytic decisions might result in different interpretations of empirical data, possibly leading to conflicting conclusions.
Failing to constrain this freedom -- experimenting with alternative analytic paths and selectively reporting findings -- inflates the chance of false discovery~\cite{simmons2011}.
Even well-intentioned experts produce large variations in analysis outcomes~\cite{silberzahn2018}, suggesting a degree of arbitrariness in analytic decisions.

In response, we investigate decision making within \emph{end-to-end quantitative analysis}: the full lifecycle of quantitative data analysis including phases of data collection, wrangling, modeling, and evaluation.
We conduct semi-structured interviews with authors of nine published studies in HCI and other scientific domains.
We pore over participants' manuscripts and analysis scripts to assess their decisions, and ask them to recall, brainstorm, and compare alternatives in every analytic step.

In this paper, we contribute the results and analysis of these interviews.
We present a visualization design for representing analytical decisions, both to communicate our interview results and as a tool for mapping future studies.
We identify recurring rationales for analytic decisions, highlighting conflicts and implicit trade-offs among options.
Next we examine the motivations for carrying out alternative analyses, a practice that exercises freedom in analytic decisions.
We subsequently discuss how participants choose what to include in research reports if they have explored multiple paths.
Finally, based on our observations, we identify design opportunities for strengthening end-to-end analysis, for instance via tracking and meta-analysis of multiple decision paths.
Given the HCI community's demonstrated interest in quantitative empirical research, we hope our findings will help inspire the design of both improved analysis tools and community standards.

\section{Related Work}

Our work is motivated by the replicability crisis and issues of ``researcher degrees of freedom.''
Our visualizations draw on the scientific workflow literature, and our interview results relate to both provenance tracking and multiverse analysis.

\subsection{Practices for Improving Replicability}
Replicability concerns have prompted scientists to re-examine how data analysis practices might lead to spurious findings.
Simmons \etal~\cite{simmons2011} describe how \emph{researcher degrees of freedom} -- the flexibility in making analytic decisions -- might inflate false-positive rates (\ie \textit{p-hacking}~\cite{nelson2018}). 
Machine learning researchers note similar issues, for example tuning random seeds can drastically alter results~\cite{henderson2018}.
Gelman \& Loken~\cite{gelman2013garden, gelman2014} argue that p-hacking need not be intentional, as \textit{implicit} decisions present similar threats.
They use a metaphor of a \emph{garden of forking paths}, with each path potentially leading to different outcomes.
Failing to address this flexibility gives rise to issues such as multiple comparison problem (MCP)~\cite{forstmeier2011,forstmeier2017,zgraggen2018}, hypothesizing after the results are known (\emph{HARKing})~\cite{kerr1998}, and overfitting~\cite{pu2018}.
As indicated by a survey of 2,000 psychologists~\cite{john2012}, p-hacking is unfortunately prevalent.

In response, scholars have endorsed a number practices, including pre-registration~\cite{cockburn2018, vantVeer2016, wagenmakers2012}, using estimation instead of dichotomous testing~\cite{anderson2001, cumming2016, dragicevic2016}, adopting Bayesian statistics~\cite{gelman2012, kay2016}, and increasing transparency in reporting~\cite{feger2019, munafo2017, nelson2018, transparent-stats}.
Wicherts \etal~\cite{wicherts2016} develop a comprehensive decision checklist for study design, data collection, analysis, and reporting.
The HCI community has also contributed empirical studies~\cite{feger2019, kale2019}, tools~\cite{eiselmayer2019, feger2019gamification, mackay2007, wacharamanotham2015} and design spaces~\cite{pu2018} for improving reproducibility.
Closest to our work are the interview studies by Kale \etal~\cite{kale2019} and Liu \etal~\cite{liu2019-alt}.
We corroborate Kale's findings on analytic decision-making strategies and Liu's observations on motivations for pursuing alternatives.
By richly diagramming our participants' analyses, we further observe recurring patterns in analysis processes, such as feedback loops and fixations.
In addition, by closely examining specific, published analyses, we identify conflicts between decision rationales and opportunism.

One perspective is that flexibility is unavoidable~\cite{silberzahn2018, simonsohn2015, steegen2016}, as well-intentioned experts may produce divergent outcomes.
In a crowdsourced study~\cite{silberzahn2018}, 29 teams analyzed the same dataset to answer the same question, yet the analyses and conclusions differ considerably.
This variation is not explained by prior beliefs, expertise, or peer-reviewed quality of the analysis~\cite{silberzahn2018}.
Fixating on a single analytic path may be less conclusive, with results dependent on arbitrary choices~\cite{simonsohn2015}.

For more comprehensive assessments, researchers have proposed \emph{multiverse analysis}~\cite{simonsohn2015, steegen2016}: evaluating all ``reasonable'' analysis specifications and interpreting results collectively (where ``reasonable'' decisions are those with firm theoretical or statistical support~\cite{simonsohn2015}).
Others have adopted multiverse analysis in practice~\cite{rae2019}, cited it as an important area for future tool work~\cite{kale2019}, designed interactive media for multiverse results~\cite{dragicevic2019}, and proposed ways to quantify multiple analysis findings~\cite{patel2015, young2017}.
Existing multiverse visualizations typically use animation~\cite{dragicevic2019}, juxtaposition~\cite{dragicevic2019, simonsohn2015}, or aggregation~\cite{steegen2016} to convey analysis outcomes; some visualize the decision space in a matrix view~\cite{simonsohn2015, steegen2016}.
Our analytic decision graph visualizations convey the multiverse decision space by depicting decisions in relation to the overall analysis process.

\subsection{Workflow and Provenance}

Prior work on computational reproducibility concerns scientific workflows~\cite{ludascher2005} -- process networks of analytic steps, that model a data analysis pipeline.
Workflow management systems (\eg~\cite{callahan2006, ludascher2005, oinn2004}) provide languages to specify workflows and record information for automation, reproducibility, and sharing~\cite{freire2008}.
These workflows are often represented as directed graphs, where nodes are computational steps and edges convey data flow.
We similarly design visualizations to communicate data analysis processes, but focus not on a singular dataflow but on the space of potential decisions.

Many workflow management systems also record provenance information, namely the history of execution steps and the environment, such as input data and parameters~\cite{freire2008}.
To visualize provenance relationships, prior work predominantly uses network diagrams~\cite{callahan2006, cheung2006, myers2003, pimentel2015, zhao2004}.
For example, VisTrails~\cite{callahan2006} visualizes provenance as a tree where a node denotes a separate dataflow that differs from its parent and an edge records the changes. 
Recent work also explores human-centered interactions with history~\cite{guo2012, kery2018}, for example supporting annotations on automatically collected provenance~\cite{guo2012} and providing lightweight interactions within Jupyter Notebooks~\cite{kery2018}.
However, understanding how analytic decisions affect outcomes is still difficult in existing provenance tools, partly because history interactions are disconnected from the analysis pipeline.
Our analytic decision graphs capture all paths taken and might serve as a navigation overview to explore history data.



\section{Methods}

To better understand decision-making in end-to-end quantitative data analysis, we conducted semi-structured interviews with authors of nine published studies. We first inspected the papers and analysis scripts, then engaged researchers in discussion about their decision rationales and possible alternatives.

\tableCodes
\subsection{Participants}

We interviewed 9 academic scientists (3 females, 6 males, age 24--72), including 6 Ph.D. students, 2 research scientists and 1 tenured professor.
Our interviewee's research fields include Human-Computer Interaction (5), Proteomics (2), Marine Biology (1), and Geography (1).
Participants' analyses cover a spectrum from directed question-answering to open-ended exploration.
P1-5 conducted confirmatory analyses: they designed controlled experiments to answer predefined research questions.
P6 explored their data to develop a biological assay.
P7 and P8 performed exploratory data analyses (EDA).
P9 gathered insights from EDA to form a hypothesis for a subsequent confirmatory experiment.

We recruited interviewees by advertising in multiple HCI and data science mailing lists.
We also identified 15 local authors from the CHI 2018 proceedings and emailed them directly, netting three participants.
Regardless of recruitment method, all interested participants filled out a survey to provide a publication and the accompanying analysis scripts.
We recruited every respondent whose publication involved quantitative data analysis and had been published in a peer-reviewed venue.

\subsection{Interview Procedure}

We interviewed one researcher at a time for 60--90 minutes.
We began each interview with an introduction describing the purpose of the interview: to understand decision making during data analysis and to collect use case examples for developing prototype tools for robust data analysis.
We then proceeded with our discussion protocol, which consisted of three phases.
The discussion focused specifically on the analysis project provided to us by the participant in the signup survey.
Afterwards, all participants were compensated with a \$20.00 gift card.

\textbf{Phase 1: Recall.} 
We first asked participants to freely propose different, yet justifiable analytic decisions. 
We encouraged participants to recall alternative paths they had considered and executed, and those raised by reviewers. 
We did this prior to other phases to elicit responses without biasing participants.

\textbf{Phase 2: Brainstorm.}
We asked participants to brainstorm alternatives using a checklist (Figure supp. 1)
based loosely on the work of Wicherts \etal~\cite{wicherts2016}.
The checklist contains common analytic decisions across stages of a typical data analysis pipeline, from data collection and wrangling to modeling and inference.
We used the checklist to help participants systematically examine all steps in the end-to-end pipeline.

\textbf{Phase 3: Compare.}
To raise options overlooked by participants in the previous phases, we discussed additional decisions we had prepared before the interview.
We generated alternative analytic proposals by perusing the paper, appendix, and analysis scripts, while consulting the checklist to ensure a comprehensive coverage of different phases of the analysis.


\subsection{Analysis of Interview Data}

All interviews were audio recorded and transcribed verbatim.
The first author analyzed the data, with iterative feedback from other authors throughout the analysis process.
As our findings might put participants in a vulnerable position, we have replaced identifiable information in figures and quotes.
For example, we might replace an identifiable variable name (\textit{autophagy substrate}) with a generic name (\textit{IV}).

We first sought to understand the overall analysis process.
From the interview data, we extracted analytic steps and their relationships to re-construct both decision points and data flow.
We drew graphs to aid interpretation, and soon realized that the graphs had greater utility beyond summarizing interview results.
We thus conducted a dedicated design exercise by outlining design goals, iterating over visual encodings, and producing visualizations, as detailed in the next section.

Next we investigated how participants made analytic decisions. 
We began by using open coding~\cite{creswell2017} as a preliminary step to identify recurring themes. 
Three themes emerged: participants provided \textit{rationales} for decisions, described their experiences in \textit{executing alternative} analyses and subsequently \textit{selective reporting} of the results.
We integrated raw codes within each theme to extract common concepts and patterns.
Table \ref{table:code} summarizes the themes and categories, along with example quotes (the quotes were edited for brevity and clarity; full quotes and relevant contexts are in later sections).
The table also lists the prevalence of each category, computed as the ratio of unique instances within each theme.
We discuss our empirical findings in the section \textit{Interview Results}.

\subsection{Limitations}

One limitation is our convenience sampling approach, which introduces potential bias.
For example, our sample is mostly composed of HCI and junior researchers.
To be clear, our research goal is to characterize the space of analytic processes and decisions, \textit{not} to quantify the prevalence of any specific activity.
Also, while our study reaches saturation in some regards~\cite{guest2006} as the last two participants did not surface new categories, our convenience sample might miss known phenomena.
Some practices currently gaining adherents, such as pre-registration followed by exploratory analysis on collected data and planned analysis based on simulated data, are not observed.
A future taxonomy might better delineate the distinction between a-priori and a-posteriori decisions.


We note violations of methodological validity when perusing participants' analyses to flag potentially problematic practices, but our judgments are subjective.
Some methods we endorse are not universally accepted, such as multiple comparison correction~\cite{cribbie2017}.
Other than methodological validity, we interpret from the perspectives of the participants as much as we could.

Participants might withhold information on potentially problematic practices.
Where possible, we complement the transcripts with what we found from participants' analysis scripts (\eg evidence of implementing multiple model formulae in R code), but not all scripts retain the full history.
Thus, there are likely additional explorations of alternatives that we are unable to observe.
In addition, all accounts of analytic decisions were given post-hoc.
Future studies are needed to inspect researchers' decision making process during the analysis event.

\section{Analytic Decision Graphs}
\label{sec:diagram}

\figurePOne

To represent participants' process and decisions, we created visualizations that we call \emph{Analytic Decision Graphs} (ADGs).
We developed ADGs in conjunction with our analysis of the transcribed interviews. We present the design of ADGs here first, so that we can refer to them in our later discussions.

\subsection{Design Goals}

ADGs aim to visualize analytic decisions in the context of end-to-end analysis pipelines.
We expect ADGs to afford two utilities.
First, with ADGs as visual illustrations, authors should be able to communicate their decisions and processes more easily.
Second, ADGs might prompt reflection on decisions, potentially encouraging consideration of further alternatives.

To review an analysis decision process, users will need to perform at least the following tasks:
\begin{itemize}[noitemsep,topsep=-4pt]
	\item Gain an \textit{overview} of the high-level analytic components.
	\item Understand the analytic \textit{steps} and their relationships.
	\item Examine and evaluate the \textit{decisions} made in each step.
\end{itemize}

From these tasks we can distill some design requirements:
\begin{itemize}[noitemsep,topsep=-4pt]
	\item \emph{Represent the input and the outcomes.} To provide context, ADGs should include inputs such as data sources and outcomes such as deliverables supporting the conclusion.
	\item \emph{Display granularity of analysis components.} ADGs should visualize both high-level modules and individual decisions.
	\item \emph{Represent relationships between the steps.} ADGs should capture various types of relationships, such as order and dependency, to organize steps into a coherent process.
	\item \emph{Visualize the rationales and the ramifications of a decision.} Visualizing rationales might help authors identify weak spots and help readers gauge the validity.
\end{itemize}

\subsection{Visual Encodings}
To meet these requirements we iterated over several designs.
We discuss the tradeoffs made and present the final design.
As ADGs should visualize both steps and their relationships, a graph is a natural representation.
We use a \textit{node} \legend{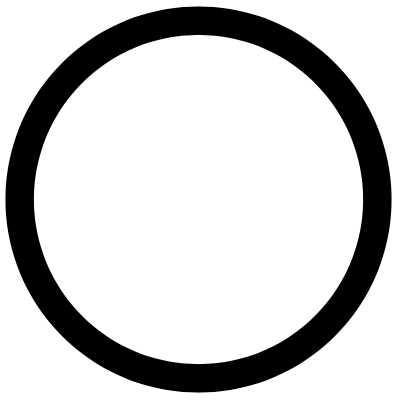} to encode a \textit{decision point} and an \textit{edge} to encode the \textit{relationship} between two decision points.
We further include auxiliary nodes with distinct shapes: rectangles \legend{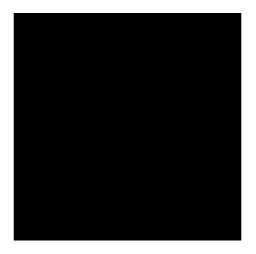} represent analysis outcomes, whereas solid dots \legend{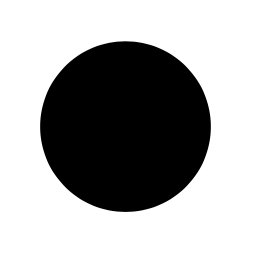} are ``dummy'' nodes. 
In an earlier design, we visualized all potential alternative choices one could make in addition to the decision point, but the graph soon grew cluttered.
We thus omit individual alternatives.

Various types of relationships exist between two decision points.
The first type is a \textit{dataflow dependency} \legend{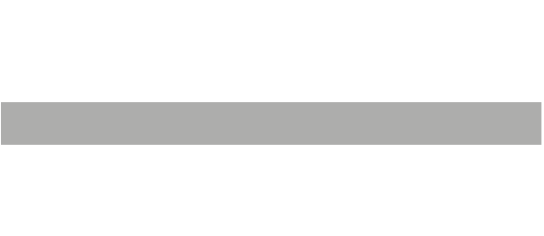}, where the output of one node is the input to another.
The second type is a \textit{procedural dependency} \legend{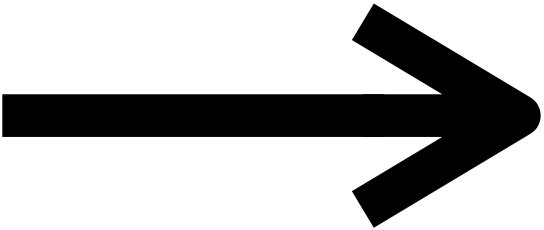}, where the downstream decision would not exist if some alternative in the upstream decision were chosen.
For example, if a researcher had chosen a frequentist model instead of a Bayesian model, she would not need to decide among different priors.
The third type is an \textit{information dependency} \legend{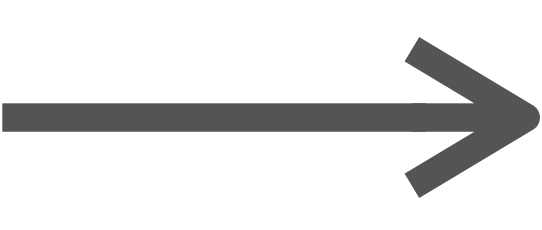}, where one decision informs another.
For example, insights from exploratory analysis might inform the hypothesis of a subsequent confirmatory experiment.
We also have \textit{feedback loops} \legend{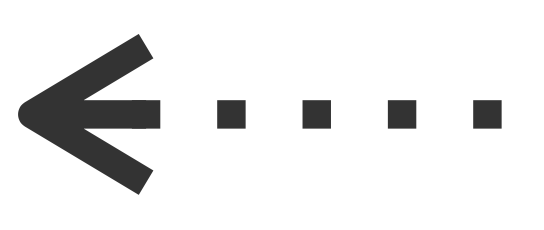}, as researchers revise an upstream decision based on the results from a downstream step.
All of these relationships appear as edges of different textures.
We further arrange the nodes vertically according to their order in the dataflow, with the top being the start.
Yet another type of relationship exists -- \textit{temporal order} -- as some decisions are made earlier than others.
We overload the vertical axis to represent temporal order when it does not conflict with dataflow dependency.

\figureDiagrams

We use a categorical color palette to represent type of decision rationale.
To reduce visual complexity, we simplify the categories of Table \ref{table:code} to three groups.
We use a red color for \textit{desired results} \legend{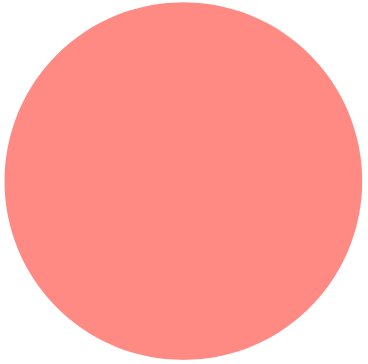} to call out potentially problematic practice; this is when researchers made the decision by weighing end results, for example discarding options that produced non-significant results.
We assign blue to \textit{data, methodology, and prior work} \legend{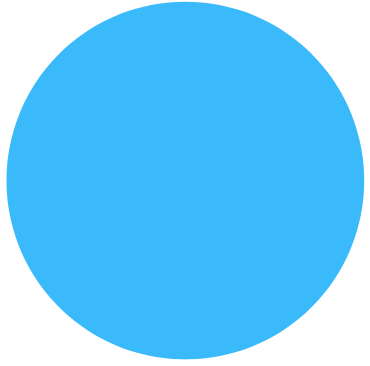}, which are relatively primary concerns.
The rest of the rationales, denoted \textit{other rationales} \legend{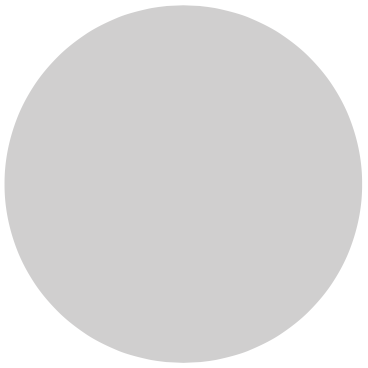}, receive a desaturated gray color.
Finally, as we (the interviewers) might propose alternatives that the participant had not thought of, we use white \legend{node} to indicate additional decisions not considered by the participants at the time of analysis.
Further information on color assignment is in the supplemental material.

The size of a node corresponds to the number of enumerated alternatives for the decision point.
The thickness of a dataflow edge conveys the number of accumulated alternatives, namely all possible combinations of alternatives of previous decisions leading to that point.
Since the accumulated total grows exponentially, we use a logarithmic scale for edge thickness.

\section{Analysis of Analytic Decision Graphs}
\label{sec:diagram_result}

We created an ADG for each participant, as shown in \rfig{fig:diagrams} (full-size diagrams are available in supplemental material).
We first describe P1's ADG (\rfig{fig:p1}) in detail, then summarize recurring patterns drawn from the ADGs for all participants.
As comparing unrelated studies is not a design goal of ADGs, care should be taken when interpreting apparent differences between graphs.
Some visual properties (\eg those described below) are meaningful to compare, but other visual differences (\eg horizontal position of nodes, edge curvature) are not.

\subsection{ADG Walkthrough for P1}

P1 designed a controlled experiment to investigate the impact of web design on reading performance.
She followed a typical confirmatory pipeline:
she operationalized (\ie defined the measurements of)  the variables germane to her research questions, collected and processed the data, built a statistical model, and interpreted the results, ultimately producing a bar chart of effect sizes with uncertainty intervals and several p-values.

The dataflow edges funnel into two linear paths leading to the end results, as opposed to a typical exploratory analysis (\eg \rfig{fig:diagrams}f) where the dataflow forks into multiple branches.
Still, P1's analysis has many feedback loops: P1 revised her analytic decisions at several steps, based on observed data, end results, and reviewer feedback.
Despite being a relatively simple pipeline with 9 decision points, P1's analysis gives rise to over 5,000 possible ways to compute the final p-values, as indicated by the width of the dataflow edge into the final node \textit{reject null hypothesis}.
Judging by the size and color of decision nodes, P1 examined model specification options thoroughly (indicated by the size of the \textit{specify model formula} node), but she appeared to place less emphasis on \textit{inference} decisions (indicated by empty nodes in the inference section).

\subsection{Summary of ADG Patterns}

Using the interpretation approach above, we analyzed ADGs for all participants.
Here are a few recurring observations.

Feedback loops are present in all analysis processes of our participants, regardless of whether the analysis is confirmatory or exploratory (\rfig{fig:diagrams}, dotted edges).
We further examine these iterative fine-tuning behaviors in the next section.

Participants often fixate on a few prominent steps while ignoring decisions in the end-to-end pipeline.
Among our participants, we observe that data and inference decisions are often neglected (\rfig{fig:diagrams}, empty nodes).
When prompted by the interview checklist or the interviewer, participants revealed that they did not recognize these steps as decision points and implicitly chose a single viable option.
On the other hand, \textit{choosing variables}, \textit{choosing models} and \textit{specifying model formula} are often considered thoroughly (\rfig{fig:diagrams}, large nodes).

Procedural branches are rare among our participants.
P1's process includes one procedural edge and no procedural branches (\rfig{fig:p1}, thick black edges); she could have considered ways to operationalize other candidate dependent variables.
The lack of such branches implies a relatively linear process where decisions were made in order, one step at a time.

Across participants, the ``multiverse'' size ranges from 16 to over 25,000,000 (median 1,632; see \rfig{fig:diagrams}, thickness of dataflow edges immediately before rectangular nodes).
We revisit issues related to scale in the discussion section.

\section{Interview Results}\label{sec:findings}

We now describe the patterns that emerged from the qualitative analysis of our interview data, following the organization of themes and categories in Table~\ref{table:code}.


\subsection{Rationales for Analytic Decisions}
\label{sec:rationale}

When participants recognized an analytic step as a decision point, they might \textit{reason} about it, identifying and evaluating options before selecting a path along which to proceed.
From 190 such instances, we identified six categories of rationales for analytic decision making.

\subsubsection{Methodology}

Methodological concerns comprised a major set of rationales (48 instances,  9/9 participants).
These arguments typically involved statistical validity, study design, and research scope.

Many methodological concerns (19 instances, 9/9 participants) were rooted in statistical validity.
Meeting model assumptions was a concern for seven participants, as they chose the statistical model best suited for the data distribution, or wrangled the inputs to satisfy model assumptions.
Participants used various strategies for the later approach: they might balance the datasets, normalize the inputs, log-transform a variable, or remove collinear variables.
Besides model assumptions, five participants supported their decision with logical arguments, pointing out mathematical properties or explaining the intuitions behind customized methods.
As a simpler example, P6 explained why she used a less common log-transformation, $log(x+1)$, to process the data: \qi{because I have a lot of zeros.} \seefig{e}{transform data}.


Validity concerns also stem from study design (21 instances, 9/9 participants).
Five participants argued that confounders were controlled for and thus were excluded in model specifications.
Four participants stressed that variables in their models strictly followed the factors, levels, and measures in their experimental design.
Participants followed a preselected plan akin to pre-registration~\cite{cockburn2018, vantVeer2016, wagenmakers2012}, though none of the studies was officially pre-registered.

Other than validity, a few rationales (8 instances, 2/9 participants) are rooted in scope, as researchers discarded alternatives outside the scope of their current research questions.
As P2 argued \seefig{a}{design typing task}: \qi{the intention of this research is to evaluate text entry, the real-life text entry. So we'll not type random text.}

\subsubsection{Prior Work}
Another group of rationales were anchored in prior work~(62 instances,  9/9 participants).
Here, we use \textit{prior work} to refer to prior studies, standard practices, and internalized knowledge.

All participants cited prior studies to support their decisions.
Besides utilizing knowledge from prior studies to inform decisions, three participants mimicked configurations from a prior work.
While this enables direct comparison with previous findings, participants might admit that alternatives warranting further considerations might exist. 
For example, P8 stated \seefig{h}{choose goodness of fit metrics}:

\q{\dots [the chosen method] is what multiple other papers have used. But there would be alternatives and we have a whole host of other model performance metrics.}

Without citing specific sources, seven participants drew on knowledge that likely resulted from a combination of prior studies, consensus, and training.
A participant referred to field consensus in outlier removal:
\qi{We did not remove any outliers. Because in the autism field, why it's called Autism Spectrum Disorder, because other Autism are considered outliers.}

Six participants in 22 instances honored \qi{standard practice}, \qi{tradition}, and \qi{convention}, sometimes without questioning its validity.
For instance, a participant followed a \qi{rule-of-thumb} of recruiting $\sim$20 participants for an experiment, though the study might be under-powered and so fail to resolve effects of smaller size \seefig{a}{choose sample size}.
A participant chose to \qi{start with a t-test, because it's standard} \seefig{b}{choose model}, though the data violated normality assumptions.
Two participants admitted that standard practices might not be best practices, but they were concerned about social aspects.
They believed that readers would accept standard practice more readily and \qi{reviewers would have asked for it.}

Five participants expressed how the lack of theory prevented them from choosing statistically valid alternatives.
Two participants avoided interaction patterns that they \qi{didn't have a strong hypothesis to include} \seefig{d}{specify model formula}.
P2 explained how tweaking alpha, a parameter in a metric to operationalize a variable, might allow one to obtain desirable outcomes, and argued against such practices because \qi{there's no reasonable theory or rationale underlying that alpha.}\seefig{a}{operationalize adjusted WPM}.
%

\subsubsection{Data}
Data constraints represented another major group of rationales~(39 instances, 8/9 participants).
Researchers were constrained by data availability, quality, and size.

Some data constraints were hard constraints.
Unavailable data might prevent participants from investigating additional variables.
P8 originally identified 23 relevant predictors from prior work, but later dropped 7 of them for which he was unable to obtain sufficient data \seefig{h}{choose predictors}.
Three participants stated that collecting more data was too costly or infeasible, as P4 complained:
\qi{we set the target beforehand, but we couldn't achieve the target group. We just tried to recruit as many groups as possible.}
\seefig{c}{choose sample size}.
Sticking with a small sample size, two participants noted that certain modeling approaches, for instance time series analysis, were infeasible.

On the other hand, some data issues allowed more room for flexibility.
What constituted clean data might be subjective, but three participants excluded noisy data at the expense of study design, for instance dropping an entire variable.
Similarly, three participants altered study designs, such as pooling variable levels, in order to achieve a larger sample size.

\subsubsection{Expertise}
Researchers also felt limited by expertise (23 instances, 8/9 participants).
They might not know what alternatives were possible, as P9 commented \seefig{i}{choose model}:
\q{And there is almost certainly some other way to do that, but I'm not sure that I would know what it is.}
When researchers had a rough notion of viable alternatives, they opted not to pursue an unfamiliar method.
P4 echoed sentiments of three participants about Bayesian analysis:
\qi{I heard something about Bayesian statistics, but I don't have any background to try more than that.} \seefig{c}{choose model}.
Two researchers deferred a decision to a statistician, who they believed had better authority over the subject.


\subsubsection{Communication}
Sometimes researchers preferred an alternative that was easier to communicate (13 instances, 6/9 participants), quoting a variety of values.
Two participants preferred an \qi{interpretable} method over \qi{methods that merely produce black-box predictions} \seefig{h}{choose model}.
A participant attempted to be \qi{consistent} with the methods he used, because \qi{otherwise the readers will be confused} \seefig{b}{choose model}.
Another participant aimed for higher \textit{generalizability} by targeting for practical use cases \seefig{a}{design typing task}.
Finally, a participant just wanted to keep things \textit{simple}, avoiding \qi{more complex} options \seefig{a}{choose IV}.
Communication concerns can come at the expense of validity.
P3 chose a statistical model suboptimal for their data distribution because \qi{to make the analysis consistent across the whole study, we just stick with one statistical test.} \seefig{b}{choose model}.

\subsubsection{Sensitivity}
Finally, researchers sometimes claimed that choosing another alternative would have little impact on the results (5 instances, 4/9 participants).
Two researchers supported the claim with logic. P8 said:
\qi{in my quick mental calculation, it seemed like it wouldn't actually make a big difference.} \seefig{h}{adjust predicted probabilities}.
Others recalled from past experience that two methods tended to produce similar results.
As they did not evaluate their current situation, perceived sensitivity might differ from actual sensitivity.

\subsection{Interactions of Rationales}
\label{sec:interactions}

We observed an interplay between decision rationales, particularly in terms of which rationales tended to dominate others.
Both our own interviews and previous studies~\cite{simonsohn2015, vantVeer2016} identify \textit{methodology} and \textit{prior work} as dominant rationales that researchers primarily rely upon.
A bottom-up, exploratory approach might include \textit{data} as a dominant rationale category, as researchers develop tentative theories to account for observed phenomena.
However, in practical situations, the analysis plan supported by the dominant rationales nevertheless accommodates various constraints concerning \textit{data}, \textit{expertise} and \textit{communication}.
\textit{Sensitivity} ignores other rationales by focusing instead on the impacts of the decision.
 
These decision rationales interact with each other, often creating conflicts.
The previous section described two ways in which the dominant categories, \textit{methodology} and \textit{prior work}, are contradictory.
First, standard practices are not always best practices; by adhering to conventions, participants might adopt a statistically faulty method.
Second, a statistically valid approach might lack theoretical support, as five participants described how they avoided such situations.
The previous section also contains ample evidence of how secondary rationales constrain and override dominant concerns.
\textit{Data}, \textit{expertise} and \textit{communication} all limit the viable methods researchers choose from, as researchers prefer a method that is familiar, easy to communicate, and feasible for the current data size.
\textit{Data}-related issues also impact study design, for instance researchers might drop a noisy variable or combine multiple levels within a variable to increase sample size.


\subsection{Motivations for Executing Alternative Analyses}
\label{sec:alt}

While some researchers \textit{reasoned} about alternatives, ruled out options, and implemented a single final decision, others \textit{executed} alternative analyses.
What spurred researchers to actualize possibilities and travel multiple analytic paths?
We found 44 instances in which participants explicitly described, or we could reasonably infer, their motivations to pursue alternatives.
We then identified four categories of motivations.

\subsubsection{Opportunism}

When being opportunistic, researchers willingly explored new alternatives,  searching for desired results in the garden of forking paths (20 instances, 7/9 participants).
Such exploratory behavior comes in two forms: one might search for patterns without a hypothesis to defend, or one might actively search for a confirmation of existing hypothesis.
The first form is sensible as long as the exploratory nature is clearly acknowledged in the publications~\cite{vantVeer2016, wagenmakers2012}.
In fact, exploratory data analysis (EDA) literature often advocates an open mindset and a comprehensive exploration before focusing on pre-defined questions~\cite{alspaugh2019, battle2019}.
Participants doing EDA all demonstrated an opportunistic attitude, as P8 described:
 \q{It was like a little experiment \dots It wasn't to test any hypothesis, but it was to explore the data in a more complete way where we could actually investigate the effects that we were interested in.}

However, we also observed opportunism among participants who reported strictly confirmatory findings (\rfig{fig:p1} \& \ref{fig:diagrams}a-d, feedback loops into red nodes).
Participants tried multiple analytic options and selected a path leading to desired results.
Such endeavors might happen in the data wrangling phase, when participants qualitatively explored data distributions and avoided analytic options unlikely to produce desired outcomes.
P1 discarded a dependent variable because it failed to yield differential results across conditions \seefig{1}{choose DV}:
\q{The distributions of accuracy are similar across questions. So, instead of looking at how different conditions affect it, we use [accuracy] as another exclusion criteria.}
Others adopted a deliberate and structured search. 
P3 tried \qi{all the different combinations} of independent variables in a model specification \seefig{b}{specify model formula}:
\q{You can think of it as a cross product, we did all of them, right? \dots we have ANOVA to test the difference of accuracy with and without considering age, and with and without considering gender, and with considering both gender and age. We did all of them.}
After an exhaustive search for patterns, he selectively reported \qi{interesting findings.}
These examples of opportunism in confirmatory analysis might increase the chance of false discovery and lead to non-replicable conclusions~\cite{gelman2013garden, nelson2018, simmons2011}.

\subsubsection{Systematicity}

Voluntary exploration was not always driven by a desire to find interesting results.
Researchers could \textit{systematically} enumerate reasonable alternatives, implement them, and evaluate the outcomes based on an objective metric (4 instances, 3/9 participants).
The key evidence to help us distinguish \textit{systematicity} from \textit{opportunism} was that the evaluation metric did not hinge on anticipated conclusions; the metric was not the end result.
Two participants enumerated model specifications and chose the best one based on the goodness of fit.
P8 also ran a local multiverse analysis and used the goodness of fit to choose the best combination of two decisions (\rfig{fig:diagrams}h).

\subsubsection{Robustness}
In another type of voluntary exploration, researchers tested alternatives \textit{after} making a decision to gauge the \textit{robustness} of the outcomes (7 instances, 4/9 participants).
After the model yielded expected results, P2 implemented two redundant tests \qi{to gain an inner confidence of the metric} \seefig{a}{choose model}.
P9 applied two protein annotation methods to corroborate the same conclusion \seefig{i}{annotate proteins}: 
\q{That is just for robustness, to say, `Hey, even if you look at orthologs of proteins that in mammals and so on are EV proteins, you see the same thing.'}

\subsubsection{Contingency}
In the case of contingency, researchers had no choice.
They had to deviate from their original plans because the planned path proved to be erroneous or infeasible (13 instances, 5/9 participants).
Contingency might arise internally, as five participants ran into a dead end and retracted to an upstream analytic step.
At a filtering step, P7 initially set loose thresholds because \qi{having more data was probably better}, but the decision backfired \seefig{f}{drop low quality reads}:
\q{But two years into the project, it was realized that this [filter] produced some very anomalous results, and we went back, and for some of the subsequent analysis we went through a more stringent filtering of the data which removed some of these anomalies.}

External contingency came from reviewers, who urged researchers to revise the analysis.
P6 switched to a Fisher's exact test from a t-test:
\qi{well, the reviewer made me do it, but I'm not sure it's the best choice.} \seefig{e}{choose model}.

\subsection{Motivations for Selective Reporting}
\label{sec:report}

After researchers executed alternative analytic paths and observed multiple outcomes, they must choose which analyses to include in publications. 
We observed 52 instances in which researchers did not report all analytic paths taken.
Why did researchers report some findings but omit others?
We identified four categories of motivations underlying selective reporting.

\subsubsection{Desired Results}
Evaluating multiple options allowed researchers to view and weigh the outcomes. Unsurprisingly, the quality of the outcome was a major criterion in selecting which alternative to report.
In opportunistic exploration, researchers searched the garden of forking paths for desired results; consequently, they typically only reported the desired results and omitted findings that were non-significant, uninteresting, or incoherent to the theory they intended to support (15 instances, 7/9 participants).

A majority of participants conducting confirmatory analysis (4/5) omitted statistically non-significant results.
When multiple results proved significant, participants selected the option with stronger implications for their intended theory.
P5 tested two ways to filter the data and both produced significant results, so she chose the larger subset such that she could argue for a greater impact of the proposed mechanism \seefig{d}{use a subset}.
Some participants included non-significant results and devised further criteria for ``interesting'' findings worthy of reporting.
To P3, interesting findings meant all significant results plus unexpected null results \qi{which we thought it might be significant but it turns out not.}
He truthfully documented initially plausible hypotheses that failed an empirical test, yet his reporting strategy also includes any hypothesis that seemed plausible post-hoc -- which is a form of HARKing~\cite{kerr1998}.

Two participants conducting EDA also omitted explored analysis paths that did not corroborate the conclusions.
Only one participant comprehensively documented alternative analyses they had performed during exploration.

\subsubsection{Similar Results}
In a few cases (5 instances, 3/9 participants), researchers relied on analytic outcomes, but argued that the outcomes were similar in terms of both the actual results and their implications.
Thus, reporting one of the alternatives was deemed sufficient.
Participants did not elaborate any criteria for selecting among similar options, implying that sensitivity alone was the reason for suppressing interchangeable analysis alternatives.

\subsubsection{Correctness}
Despite having access to the analytic outcomes, sometimes participants did not utilize this information.
Instead, they fell back to using rationales described in the \textit{decision rationales} theme, most frequently \textit{methodology} and \textit{prior work}, to remove analytic approaches they considered incorrect (16 instances, 7/9 participants).
Such practices often ensued from an exploration out of contingency or robustness.
For example, researchers switched to an alternative method requested by reviewers, omitting the original, presumably flawed, method.
However, sometimes the motivation for exploring alternatives was unclear and we do not know whether the correctness argument was formed before or after seeing the results.
The latter scenario, namely coming up with post-hoc explanations for \textit{desired results}, is precisely HARKing \cite{kerr1998}.

\subsubsection{Social Constraints}
Finally, social constraints could prevent participants from reporting certain findings (16 instances, 7/9 participants).
Colleagues and reviewers might disapprove of particular analysis methods.
P2 did not report his experimental code on Bayesian analysis because his \qi{colleagues don't seem to favor that} \seefig{a}{choose model}.
P8 similarly complained that he did not have full control over reporting: 
\q{I'm a second author and many decisions made in the publication, in the manuscript writing, and figure making were decisions against my wishes.}

Two participants mentioned that reporting every detail would exceed the page limit.
In response, P3 deleted the alternative taking up more space and P2 removed a finding perceived by the authors to be \qi{not of interest.}

Researchers might voluntarily cater to communicative concerns to make figures and manuscripts easier to understand.
Two participants applied additional filtering to a visualization to reduce over-plotting; they omitted the original plot and parameters.
Several participants removed analysis methods unfamiliar to the audience.
P2 stated that describing Bayesian analysis in an accessible way would be too much work, and P9 simply claimed that a method would confuse readers.

\section{Discussion}\label{sec:discussion}

In this work, we pored over nine published studies and interviewed the authors to discuss analytic decisions in the end-to-end quantitative data analysis.
We presented common rationales for analytic decisions and discussed how researchers trade off between options.
We observed various reasons for exploring alternatives and selectively reporting results.
We also introduced Analytic Decision Graphs and discussed recurring patterns along analysis processes.
Together, these results help us better understand current practices in the midst of the replicability crisis and how we might start to revise them.
Below, we discuss design opportunities for supporting users in making and communicating analytic decisions.

\subsection{Analysis Diagramming \& Provenance Tracking}

In many instances our respondents were limited in coming up with alternatives: they might fail to recognize an analytic step as a decision (\eg following default settings), adopt a single option without considering alternatives (\eg making the same decision as a previous study), or overlook possible alternatives due to \textit{expertise}.
A corresponding avenue for future research concerns analysis \emph{linters} or \emph{recommenders}, in which tools flag potentially problematic practices (such as the feedback loops observed in our interviews), recommend alternative methods, or even automatically suggest a preferred method based on statistical validity~\cite{cashman2019, jun2019, wacharamanotham2015}.
One strategy for such tools is to enable higher-level specifications of analysis goals (\eg specifying annotated model inputs and outputs rather than explicit test types or formulae), from which appropriate analysis methods might be synthesized in conjunction with the data~\cite{jun2019}.
Another strategy is to leverage the abundance of online analysis code~\cite{rule2018} to mine patterns of decisions and alternatives, which might be useful for building automatic recommenders.

In some cases, our respondents evaluated multiple alternatives and then engaged in selective reporting.
 Integrating diagramming methods with provenance tracking could provide some level of automated documentation, for example by analyzing executed code paths to model and visualize the various alternatives that were explored (\cf \cite{kery2019}). Similar elicitation and tracking strategies have also been suggested for reducing false discovery during exploratory visualization~\cite{zgraggen2018}.

Even with complete documentation of analysis history, hindsight bias might lead researchers to unintentionally misremember \emph{post hoc} explanations developed after conducting analysis as motivating \emph{a priori} hypotheses~\cite{kerr1998}.
Tools for mapping analysis decisions might promote more comprehensive assessment \emph{a priori}. 
By instantiating decision points and providing analytic checklists~\cite{wicherts2016}, analysis tools might do more to promote \emph{planning}, not just \emph{implementation}.
For example, an analysis team might manually author, annotate, and debate an analytic decision graph and corresponding rationales \emph{a priori}.
The results could then document and aid communication of decision points and rationales. 
Overviews of the end-to-end analysis process could also guide implementation work, for example with decision graph nodes linked to corresponding analysis code snippets (\ie cells in a computational notebook).

\subsection{Multiverse Specification \& Analysis}

While the above methods focus on documenting decisions and selecting a preferred path, many ``reasonable'' alternatives may exist. Proponents of \emph{multiverse analysis}~\cite{simonsohn2015, steegen2016} have argued for preserving such decisions and evaluating them collectively. However, the design and evaluation of tools for both specifying and evaluating multiverse analyses remains an open challenge.

Authoring a multiverse analysis may be tedious, as analysts have to write scripts to manually execute all possible combinations of reasonable alternatives. Future tools could provide better scaffolding for defining decision points and procedural branches without devolving into a morass of multiple, largely redundant analysis scripts~\cite{guo2012}. Inspiration might be taken from design tools for parallel prototyping~\cite{hartmann2008, terry2004}.

Second, interpreting the outcomes of a vast number of analyses is difficult.
Visualizations that juxtapose or animate individual outcomes~\cite{dragicevic2019} may not scale, and may fail to accurately convey the relative sensitivity of decision points. In addition, some of our participants bypassed decision making if they perceived the \textit{sensitivity} to be low; they did not always verify if the decision indeed had limited influence on the results.
Future tools might aggregate subsets of outcomes, and quantify the end-to-end statistical variance via a meta-analysis of multiverse results~\cite{patel2015, young2017}.
Multiverse analysis tools might assess sensitivity across decision points and identify high-impact decisions for further consideration.

Finally, multiverse analysis also poses a number of underlying systems challenges. How might one optimize multiverse evaluation, for example by efficiently reusing shared computation across ``universes,'' or by using adaptive sampling methods to more efficiently explore a parameter space?

\subsection{Sociotechnical Concerns}

While new analysis tools might help improve systematic consideration and communication of analysis alternatives, they must operate within an accepting social environment.
We are hardly the first to note that the urges to ``tell a good story,'' sidestep unfamiliar methods, and appease reviewers can undermine a full and accurate accounting of one's research~\cite{kerr1998}, and our interviews confirm their persistence.
If publication incentives and reviewer criteria remain unchanged, a provenance tracking tool that reveals problematic choices, or multiverse tools that produce more comprehensive yet more complex and unfamiliar outputs, may be abandoned in favor of the status quo.
Accordingly, improving the reliability of end-to-end analysis must also be a community priority, ranging from the standards and practices of peer review to
how we educate researchers, new and old.
We hope that the decision making and selective reporting rationales identified in our interview analysis provide useful insights for the design of both improved analysis tools \emph{and} community processes.

\section{Acknowledgements}
We thank our participants, the anonymous reviewers (especially the shepherd), Alex Kale, Eunice Jun, Rene Just, Tongshuang Wu, and IDL members for their help. 
This work was supported by a Moore Foundation Data-Driven Discovery Investigator Award and NSF Award 1901386.

\section{Supplemental Material}
Additional supporting information on methods and graphs may be found at https://osf.io/m5cph/.

\balance{}

\bibliographystyle{SIGCHI-Reference-Format}
\bibliography{multiverse}

\end{document}